\documentclass[11pt,english]{article}
\usepackage[left=1in,right=1in,top=1in,bottom=1in]{geometry}
\pdfoutput=1

\usepackage[table]{xcolor}
\usepackage[utf8]{inputenc}
\usepackage{amsfonts}
\usepackage{ragged2e}
\usepackage{amsmath}
\usepackage{amssymb}
\usepackage{graphicx}
\usepackage{wrapfig}
\usepackage{physics}
\usepackage{hyperref}
\usepackage{ragged2e}
\usepackage{float}
\usepackage{authblk}
\usepackage{caption}
\usepackage{subcaption}
\usepackage{csquotes}
\usepackage{appendix}

\usepackage[
backend=bibtex,
maxbibnames=10,
style=ieee
]{biblatex}

\numberwithin{equation}{section}
\numberwithin{table}{section}
\numberwithin{figure}{section}

\author{
T.~Goda,$^1$, K.~Kutak$^{1,2}$ and S.~Sapeta$^1$\\\,\\
$^1$ 
{\small\it The H.\ Niewodnicza\'nski Institute of Nuclear Physics PAN,}\\ 
{\small\it Radzikowskiego 152, 31-342 Krak\'ow, Poland}\\\,\\
$^2$ 
{\small\it Physics Department, Brookhaven National Laboratory,}\\
{\small\it Upton, NY 11973, USA}\\
}	 
\title{Sudakov effects and the dipole amplitude}

\date{\today}

\begin{document}

\maketitle

%--------------- preprint numbers ---------------------
\vspace{-25em}
\begin{flushright}
  IFJPAN-IV-2022-16\\
\end{flushright}
\vspace{20em}
%-----------------------------------------------------------------

\begin{abstract}
In this study we incorporate the Sudakov form factor into the dipole factorization formula, where the hard scale of the former is provided by the photon virtuality $Q^2$.
We obtain a general formula which we then apply to the well-known GBW and BGK saturation models.
Parameters of the above Sudakov-improved models are successfully fitted to the $F_2$ data from HERA. We observe, in particular, that inclusion of the Sudakov factor on top of the GBW model improves description of data at large $Q^2$.      
\end{abstract}
%%%%%%%%%%%%%%%%%%%%%%%%%%%%%%%%%%%%%%%%%%%%%%%%%%%%%%%%%%%%%%%%%%%%%%%%%%%%%%%%%%%%%%%%%%5
\section{Introduction}
%%%%%%%%%%%%%%%%%%%%%%%%%%%%%%%%%%%%%%%%%%%%%%%%%%%%%%%%%%%%%%%%%%%%%%%%%%%%%%%%%%%%%%%%%%

Deep Inelastic Scattering (DIS) processes play central role in studies of the structure of hadrons and nuclei~\cite{roberts_1990,collins_2011}.  For many years, the HERA collider, located at DESY, remained the chief experimental facility in that area of research and lead to, among other things, very precise results for the $F_2$ structure function~\cite{H1:2009pze,H1:2015ubc}.
With the prospect of the Electron Ion Collider~(EIC)~\cite{NAP25171}, DIS remains to be of great interest for further studies.

For kinematic configurations in which a hadron is probed at small longitudinal momentum fractions, $x$, it appears as being build up by a large number of soft gluons.
Therefore, one is interested in distributions of those gluons over the collinear momentum fraction, $x$, and the transverse momentum, $k_t$. 
Such functions are called Transverse Momentum Dependent Parton Distribution Functions~(TMDs) \cite{Angeles-Martinez:2015sea} and are closely related to the color-dipole amplitude/cross section. 

As is well known, the Balitsky--Fadin--Kuraev--Lipatov (BFKL) equation~\cite{Balitsky:1978ic, Kuraev:1977fs} predicts a strong rise of gluon distributions with decreasing $x$, which grows like $x^{-\alpha_P+1}$, where $\alpha_P-1=\frac{4 \alpha_s N_c}{\pi}\log2$, and eventually violates unitarity~\cite{Gribov:1981ac,Gribov:1983ivg,Mueller:1985wy}. 
To cure the problem one accounts  for higher order corrections \cite{Fadin:1998py} and performs resummations \cite{Kwiecinski:1997ee,Salam:1998tj,Ciafaloni:2003kd,Ciafaloni:2003rd,Ciafaloni:2003ek,Hentschinski:2012kr,Li:2022avs}. However, while these  corrections slow down the growth of gluons, they do not tame it. In particular, the gluon density diverges for small values of the transverse momentum. Thus, one needs to take into account also recombinations of gluons leading to saturation~\cite{Gribov:1983ivg,Mueller:1985wy} (see \cite{Kovchegov:2012mbw} for review) that is modelled via nonlinear evolution equations, such as the Balitsky-Kovchegov~ (BK)~\cite{Balitsky:1995ub,Kovchegov:1999yj} and Jalilian-Marian--Iancu--McLerran--Weigert--Leonidov--Kovner (JIMWLK)~\cite{Kovner:1999bj,Kovner:2000pt,Iancu:2000hn,JalilianMarian:1997dw,JalilianMarian:1997gr,JalilianMarian:1997jx}, describing the gluonic state known as the Color Glass Condensate~\cite{McLerran:1993ni, Ayala:1995kg, Kovchegov:1996ty,Jalilian-Marian:1996mkd,Iancu:2000hn,Iancu:2001ad,Ferreiro:2001qy,Iancu:2002xk}.  
The taming of the growth is referred to as the {\it saturation phenomenon}.
Saturation can be also successfully modeled phenomenologically, with the most notable examples of the Golec-Biernat--W\"usthoff~(GBW)~\cite{Golec-Biernat:1998zce}, Bartels--Golec-Biernat--Kowalski (BGK)~\cite{Bartels:2002cj},
Kowalski--Teaney (KT)~\cite{Kowalski:2003hm}, 
McLerran--Venugopa-lan~(MV)~\cite{McLerran:1993ni}, Shaw--Kerley--Forshaw (SKF)~\cite{Forshaw:1999uf} and Iancu--Itakura--Munier (IIM)~\cite{Iancu:2003ge} models.  

In recent years, it became increasingly evident~\cite{vanHameren:2020rqt, VanHaevermaet:2020rro, Collins:1984kg, Gribov:1983ivg}, that on top of the modeling of the non-linear gluon evolution at small $x$, one should also take into account the so-called Sudakov effects, which become relevant for processes with two distinct scales, like, for example, $k_t$ and $p_T$, where the latter is a transverse momentum of final state hard object.
In such cases, one encounters large logarithms $\log(p_T^2/k^2_t)$, which should be resummed~\cite{Collins:1984kg}. 
As shown in Refs.~\cite{Mueller:2012uf, Mueller:2016gko}, simultaneous resummation of $\log(1/x)$ and $\log(p_T^2/k_t^2)$ can be achieved consistently, owing to separation of contributing regions in the phase space. 
In particular, in Ref.~\cite{Xiao:2017yya}, the Sudakov logarithms were resummed for the dipole gluon density. Furthermore, the in the context of DIS and $F_2$ structure function the Sudakov form factor 
is an element of extension of BFKL to account for coherence leading to the Catani--Ciafaloni--Fiorani--Marchesini (CCFM) evolution equation \cite{Ciafaloni:1987ur,Jung:2010si}, as well as it is a basis for Monte Carlo tools for simulations of parton branching \cite{Hautmann:2017fcj,BermudezMartinez:2018fsv,Hautmann:2022xuc}. In this case it's role is to at least partially account for DGLAP logs in the low-x frameworks.  

In this work we follow this path and  
we implement the Sudakov form factor in the context of the color-dipole cross section.
In the leading-$\log(1/x)$ approximation, the unintegrated gluon density, and equivalently the dipole cross section are hard scale independent, as the hard scale dependence is a subleading effect \cite{Kimber:1999xc, Kimber:2000bg}. 
However, in the region of moderate-$x$, one needs to consider also the dependence on $Q^2$~\cite{Kimber:1999xc, Kimber:2000bg}. The Sudakov factor described earlier introduces additional hard scale dependence into the $F_2$ structure function and one may expects that the region of applicability in $x$ will extend further into the moderate-$x$ region.
For earlier proposals for combining dipole cross section with the Sudakov factor we refer the Reader to~\cite{Motyka:2002ww}.

After a brief description of the dipole picture of DIS in Section~\ref{sec:Dipole amplitude and Sudakov}, we discuss how saturation models can be combined with the Sudakov form factor.
In Section~\ref{sec:GBW and BGK models with Sudakov}, two models on which we develop our modified versions, namely the Golec-Biernat--W\"usthoff (GBW) model~\cite{Golec-Biernat:1998zce} and the Bartels--Golec-Biernat--Kowalski (BGK) model~\cite{Bartels:2002cj}, are briefly summarized. In Section~\ref{sec:Fits of the Sudakov-improved saturation models}, the result of fitting the saturation models with the Sudakov factor, to the data from the HERA~\cite{Abt:2017nkc}, is presented.

%%%%%%%%%%%%%%%%%%%%%%%%%%%%%%%%%%%%%%%%%%%%%%%%%%%%%%%%%%%%%%%%%%%%%%%%%%%%%%%
\section{Dipole amplitude and Sudakov form factor}\label{sec:Dipole amplitude and Sudakov}
%%%%%%%%%%%%%%%%%%%%%%%%%%%%%%%%%%%%%%%%%%%%%%%%%%%%%%%%%%%%%%%%%%%%%%%%%%%%%%

In the dipole picture of DIS, a photon with virtuality $Q^2$ fluctuates into a pair of quark and antiquark, with momentum fractions $z$ and $(1-z)$, respectively, long before the interaction with the nucleon. 
The cross section of inclusive DIS involving the photon with polarization $P$, at leading order~(LO) factorizes as~\cite{Golec-Biernat:1998zce, Kovchegov:2012mbw}
\begin{equation}
    \sigma^{\gamma^* p}_{P}(x,Q^2)=\int d^2 \mathbf{r} \int^1_0 dz |\Psi_{P}(z,\mathbf{r},Q^2)|^2 \sigma_{\mathrm{dipole}}(x,r),
    \label{eq:factorization}
\end{equation}
where the photon wave function, $\Psi_P(z, \mathbf{r},Q^2) $, describes the decay of a virtual photon into a quark-antiquark pair, separated by the dipole size $r$, and the dipole cross section, $\sigma_{\mathrm{dipole}}(x,r)$, describes the interaction of the quark/antiquark with the target\footnote{For the result at NLO accuracy see~\cite{Beuf:2017bpd}.}. The dipole cross section is an integral of the dipole amplitude over the impact parameter $b$
\cite{Kowalski:2003hm}
\begin{equation}
    \sigma_{\mathrm{dipole}}(x,r)=2\int d^2\mathbf{b} N(x,r,\mathbf{b})\,.
    \label{eq:impact}
\end{equation}
The dipole amplitude can be obtained by solving the aforementioned QCD evolution equations like BK~\cite{Balitsky:1995ub,Kovchegov:1999yj} or JIMWLK~\cite{Kovner:1999bj,Kovner:2000pt,Iancu:2000hn,JalilianMarian:1997dw,JalilianMarian:1997gr,JalilianMarian:1997jx}.
However, it can be also modelled, which is often useful as it allows one to gain valuable analytic insight.
(See Ref.~\cite{Kutak:2004ym} for a comparison of a solution of the BK equation and a saturation model that will be discussed in the next section.)

The dipole cross section, $\sigma_{\mathrm{dipole}}(x, r) $, and the dipole unintegrated gluon distribution, $\mathcal{F}(x,k_t^2)$, are related to each other by \cite{Golec-Biernat:1999qor,Bartels:2002cj}
\begin{align} 
    \sigma_{\mathrm{dipole}}(x,r) &= \frac{4 \pi}{N_c} \int\frac{d^2 \mathbf{k}_t }{k_t^2} \alpha_s \mathcal{F}(x, k_t^2) (1-e^{i {\mathbf{k}_t\cdot \mathbf{r}}}),
    \label{eq:ugdtosigma}\\
    \alpha_s \mathcal{F}(x,k_t^2) &= \frac{N_c }{4 \pi}\int \frac{d^2  \mathbf{r}}{(2\pi)^2} e^{i {\mathbf{k}_t\cdot \mathbf{r}}}  \nabla^2_{\mathbf{r}} \sigma_{\mathrm{dipole}}(x,r).
    \label{eq:sigmatougd}
\end{align}
By combining these expressions one obtains
\begin{equation}
    \sigma_{\mathrm{dipole}}(x,r) =\int \frac{d^2 \mathbf{k}_t}{(2\pi)^2k_t^2}\left( 1- e^{i {\mathbf{k}_t\cdot \mathbf{r}}}\right) \int d^2 \mathbf{{r'} } e^{i{\mathbf{k}_t\cdot \mathbf{r'} }}\nabla^2_{\mathbf{{r'} }}\sigma_{\mathrm{dipole}}(x,{r'} ).
    \label{eq:dipoletodipole}
\end{equation}
The Sudakov form factor, $S(r,Q^2)$, can be included by using the above formula and generalizing it to
\cite{Xiao:2017yya}
\begin{equation}
    \Sigma_{\mathrm{dipole}}(x,r,Q^2) =\int \frac{d^2 \mathbf{k}_t}{(2\pi)^2 k_t^2} \left( 1- e^{i {\mathbf{k}_t\cdot \mathbf{r}}}\right) \int d^2 \mathbf{ {r'} } e^{i{\mathbf{k}_t\cdot \mathbf{r'} }}e^{-S({r'} ,Q^2)} \nabla^2_{\mathbf{{r'} }}\sigma_{\mathrm{dipole}}(x,{r'} )\,,
\end{equation}
where the new object $\Sigma_\text{dipole}$, on left hand side, now depends on the hard scale, $Q^2$.
The integral can be partially performed analytically and one obtains
\begin{equation}
    \Sigma_{\mathrm{dipole}}(x,r,Q^2) =\int^r _0 d {r'}   {r'}  \log\left(\frac{r}{{r'} }\right) e^{-S({r'} ,Q^2)} \nabla^2_{{r'} }\sigma_{\mathrm{dipole}}(x,{r'} ).
    \label{eq:gbssigma}
\end{equation}
The above formula is general, {\it i.e.} it can be used with any dipole cross section\footnote{For other approaches to account for the Sudakov form factor, together with the dipole cross section and the nonlinear evolution, see \cite{Kutak:2011fu,Kutak:2014wga,Kutak:2012qk,Zheng:2019zul}.}.

%%%%%%%%%%%%%%%%%%%%%%%%%%%%%%%%%%%%%%%%%%%%%%%%%%%%%%%%%%%%%%%%%%%%%%%%%%%%%%%%%%%%%%%%%
\section{GBW and BGK models with Sudakov form factor}\label{sec:GBW and BGK models with Sudakov}
%%%%%%%%%%%%%%%%%%%%%%%%%%%%%%%%%%%%%%%%%%%%%%%%%%%%%%%%%%%%%%%%%%%%%%%%%%%%%%%%%%%%%%%%%
The dipole cross section of the GBW model is given by~\cite{Golec-Biernat:1998zce} 
\begin{equation}
    \sigma_{\mathrm{GBW}}(x ,r )=\sigma_{0} \left(1-e^{-\frac{r^2}{4} Q_0^2\left(\frac{x_0}{x}\right)^{\lambda} }\right),
    \label{eq:gbw}
\end{equation}
and a modified variable 
\begin{equation}
    x_m=x \left(1+\frac{4 m_f^2}{Q^2}\right),
\label{eq:modx}
\end{equation}
where $m_f$ is an effective mass of a quark $f$,
was used in place of the Bjorken scaling variable~$x$. In the rest of this paper, the subscript $m$ is dropped. Along with the above modification, effective mass of the light quarks $m_l=0.14\, \mathrm{GeV}$ was introduced in Ref.~\cite{Golec-Biernat:1998zce} in order to study the photoproduction limit of DIS~($Q^2\rightarrow0$). 

The GBW model incorporates two main features which characterize the two regions of $r$ separated by the saturation scale $Q_s(x)$, which we discuss in detail later.
That is when $r\ll 2/Q_s(x)$,
\begin{equation}
    \sigma_{\mathrm{GBW}}\simeq\frac{r^2}{4} Q_0^2\left(\frac{x_0}{x}\right)^{\lambda},
\end{equation}
while when $r\gg 2/Q_s(x)$, the dipole cross section approaches a constant value.
However, the small-$r$ limit of the dipole cross section reads~\cite{Bartels:2002cj}
\begin{equation}
    \sigma_{\mathrm{dipole}}(x,r)\simeq \frac{\pi^2 r^2\alpha_s(\mu^2) x g(x,\mu^2)}{3},
\end{equation}
where $g(x,\mu^2)$ is the integrated gluon distribution at the scale $\mu^2=\frac{C}{r^2}$.
In order to account for this behaviour, Bartels, Golec-Biernat and Kowalski proposed an improved version of the saturation model, in which the dipole cross section is written in the form~\cite{Bartels:2002cj}
\begin{equation}
    \sigma_{\mathrm{BGK}} (x,r)=\sigma_0 \left[ 1-\exp\left( - \frac{\pi^2 r^2\alpha_s(\mu^2) x g(x,\mu^2)}{3 \sigma_0} \right) \right],
    \label{eq:bgk}
\end{equation}
where $\mu^2=\frac{C}{r^2} + \mu_0^2$.
As a result, now, the small-$r$ behaviour is enhanced with the gluon distribution $x g(x, \frac{C}{r^2})$.
The integrated gluon distribution, $g(x,\mu^2)$, follows the DGLAP evolution~\cite{Golec-Biernat:2017lfv},
and thus the modification~(\ref{eq:bgk}) properly introduced the QCD behaviour in the large $Q^2\sim1/r^2$ region\footnote{The $z$-integrated integrand of Eq.~(\ref{eq:factorization}) peaks around $r\sim2/Q$ and thus the modification affects primarily the large-$Q$ region.}. 
At the same time, in the large-$r$ region, where $\mu^2\simeq \mu_0^2$, $\sigma_{\mathrm{BGK}}$ recovers the form of the GBW model as desired.  

As discussed in the previous section, the Sudakov form factor can be combined with the dipole models with help of the formula~(\ref{eq:gbssigma}).
In the present study, we focus on the leading-order, perturbative Sudakov factor~\cite{Xiao:2017yya},
\begin{equation}
    S^{(1)}_\mathrm{pert}(r,Q^2)=\frac{C_A }{2 \pi} \int^{Q^2}_{\mu_b^2  } \alpha(\mu^2 )\frac{d \mu^2}{\mu^2}  \log\left(\frac{Q^2}{\mu^2}\right).
\end{equation}
For the case of the running coupling $\alpha_s(\mu^2)=1/(b_0 \log \frac{\mu^2}{\Lambda^2_\text{QCD}} )$, we get
\begin{multline}
    S^{(1)}_\mathrm{pert}(r,Q^2) =\frac{C_A}{2 \pi b_0} \Big[-\log\left(\frac{Q^2}{\mu^2_b}\right) \\
    +\left(\frac{1+\alpha(\mu_b^2)  b_0 \log \left(\frac{Q^2}{\mu_b^2}\right) }{\alpha(\mu_b^2)   b_0}\right) \log\left( 1+\alpha(\mu_b^2)   b_0 \log \left(\frac{Q^2}{\mu_b^2}\right) \right) \Big],
    \label{eq:sudakov}
\end{multline}
where $b_0=\frac{11 C_A-2 n_f}{12}$.
The lower limit of the integral $\mu_b=C_S/r$, where $C_S=2e^{-\gamma_\mathrm{E}}$, and $\gamma_\mathrm{E} \approx 0.577$ is the Euler-Mascheroni constant.

Two points have to be clarified concerning the above expression.
Firstly, the lower limit of the integration is restricted to the region $\mu_b<Q$ such that for $\mu_b>Q$, $S(r,Q)=0$. That is to say that in the limit $Q^2\rightarrow0$, the effect of the Sudakov factor is nil. Consequently, for a fixed value of $Q^2$, the small-$r$ limit of the original dipole cross section is recovered.  Such restriction is not present in the original CSS formula~\cite{Collins:1984kg} (See also Ref.~\cite{Collins:2016hqq} for the improved treatment for small $r$.) but as we are concerned with relatively low values of $Q^2$, one needs to consider the region $1/r^2\sim k_t^2 >Q^2$.
In short, in the limit $Q\rightarrow 0$ we recover the original GBW/BGK model. Also in the limit $r \rightarrow 0$, again,  we recover those two models.
Secondly, as the dipole size $r$ becomes large, the integration enters the non-perturbative region. In this region, one can freeze $r$ by the so-called ``$b_*$-prescription'' \cite{Collins:1984kg}, in which 
\begin{equation}
    \mu_b^2=C_S^2/r_*^2= C_S^2/r^2+C_S^2/r_{\mathrm{max}}^2.
\end{equation}
The constant $r_{\mathrm{max}}$ is chosen such that in the large-$r$ region $\mu_b^2\simeq C_S^2/r_{\mathrm{max}}^2 \gg \Lambda_{\mathrm{QCD}}^2$.

In principle, one can include also the non-perturbative Sudakov factor~\cite{Collins:1984kg, Prokudin:2015ysa}:
\begin{equation}
    S(r,Q^2)= S_\mathrm{pert}(r,Q^2)+S_\mathrm{np}(r,Q^2),
\end{equation}
where~\cite{Prokudin:2015ysa}
\begin{equation}
    \label{eq:Snp}
    S_\mathrm{np}(r,Q^2)=g_1r^2+g_2\log\left(\frac{r}{r_*}\right)\log\left(\frac{Q}{Q_0}\right).
\end{equation}
However, 
considering that $\nabla^2_r \sigma$ in Eq.~(\ref{eq:gbssigma}) contains the factor $e^{-a r^2}$, where $a$ depends on the models, it is apparent that the general behaviour of $S_\mathrm{np}\sim r^2$ is already present in the GBW and BGK models. 
The non-perturbative Sudakov form factor is primarily important in the region $1/r<Q\lesssim 1/r_{\mathrm{max}}$. In this  region, the second term in Eq.~(\ref{eq:Snp}) is small compared to the first term, not only because of the suppression by $\log(Q/Q_0)$, but also because in the large-$r$ region, $r^2>\log\left({r}/{r_*}\right)$.  Therefore, one expects the contribution of the second term to be comparatively smaller than that of the $\sim r^2$ term, and the only possible region where it may contribute corresponds to not too small $Q$ and not too large $r$. One, however, needs to keep in mind that the Sudakov factor enters via the integrand of $\Sigma_\mathrm{dipole}$, Eq.~(\ref{eq:gbssigma}). That is to say that in the large-$r$ region of $\Sigma_\mathrm{dipole}$, it has effect from the small-$r'$ region of the integrand.

Indeed, as we shall briefly discuss in Section~\ref{sec:Fits of the Sudakov-improved saturation models}, an improvement by the non-perturbative Sudakov factor is less than 5\% for the BGK model and practically nil for the GBW model, and thus we neglect it in this study. Since diffractive DIS is more sensitive to the large-$r$ region, it may become necessary to include the non-perturbative factor in that case.

In our study, we shall employ, for both $g(x,\mu^2)$ and the Sudakov factor, an alternative form of $\mu_b$ proposed in Ref.~\cite{Golec-Biernat:2017lfv}:
\begin{equation}
    \mu_b^2= \frac{\mu_0^2}{1-e^{-r^2\frac{\mu_0^2}{C} }},  
    \label{eq:newstar}
\end{equation}
where $\mu_0^2=C/r_{\mathrm{max}}^2$.

%%%%%%%%%%%%%%%%%%%%%%%%%%%%%%%%%%%%%%%%%%%%%%%%%%%%%%%%%%%%%%%%%%%%%%%%%%%%%%%%%%%%%%%%%%%%%%%%%%%%%%%%%
\section{Fits of the Sudakov-improved saturation models}\label{sec:Fits of the Sudakov-improved saturation models}
%%%%%%%%%%%%%%%%%%%%%%%%%%%%%%%%%%%%%%%%%%%%%%%%%%%%%%%%%%%%%%%%%%%%%%%%%%%%%%%%%%%%%%%%%%%%%%%%%%%%%%%%%

\begin{table}[t]
    \begin{subtable}[t]{\textwidth}
        \centering
        \vspace{2mm}
\begin{tabular}{|c|c|c|c|c|}
\hline
  &  $\sigma_0 \;[\mathrm{mb}]$ &  $x_0 (10^{-4})$ &  $\lambda$ &  $\chi^2/\mathrm{dof}$ 
\\\hline
$\mathrm{GBW}$ & 19.1 & 2.58 & 0.322 & 4.44 \\ \hline
$\mathrm{GBW+Sud}$& 18.6 & 3.11 & 0.299 & 2.66 \\ \hline
\end{tabular} 
\vspace{2mm}
    \end{subtable}
    \begin{subtable}[t]{\textwidth}
        \centering
        \vspace{2mm}
\begin{tabular}{|c|c|c|c|c|c|c|}
\hline
 &  $\sigma_0\; [\mathrm{mb}]$ &  $A_{\mathrm{g}}$ &  $\lambda_{\mathrm{g}}$ &  $C$ &  $\mu_{0}^2\;[\mathrm{GeV^2}]$ &  $\chi^2/\mathrm{dof}$ 
\\\hline
$\mathrm{BGK}$ & 23.3 & 1.18 & 0.0832 & 0.329 & 1.87 & 1.56 \\ \hline
$\mathrm{BGK+Sud}$ & 22.2 & 8.67 & -0.500 & 0.670 & 3.83 & 1.21 \\ \hline
\end{tabular} 
\vspace{2mm}
    \end{subtable}
    \caption{ The parameters and $\chi^2$ per degrees of freedom of the GBW and BGK models (with and without the Sudakov factor, Eq.~(\ref{eq:sudakov})) for cases with mssless light quarks.}
    \label{table:fitresults}
    \vspace{3mm}
    \centering
    \begin{subtable}[t]{0.475\textwidth}
        \centering
        %\resizebox{0.35\textwidth}{!}{
\begin{tabular}{|c|c|c|}
\hline
$Q_{\mathrm{up}}^2 \;[\mathrm{GeV^2}]$&  GBW & GBW+Sud 
\\\hline
5 & 1.55 & 1.55 \\ \hline
25 & 1.46  & 1.41 \\ \hline
50 & 1.97  & 1.83 \\ \hline
100 & 2.36  & 2.15 \\ \hline
650 & 4.44  & 2.66 \\ \hline
\end{tabular}
%}

    \end{subtable}
    \hfill
    \begin{subtable}[t]{0.475\textwidth}
        %\resizebox{0.35\textwidth}{!}{
\begin{tabular}{|c|c|c|}
\hline
$Q_{\mathrm{up}}^2 \;[\mathrm{GeV^2}]$&  BGK & BGK+Sud
\\\hline
5 & 1.63  & 1.59 \\ \hline
25 & 1.42 & 1.30 \\ \hline
50 & 1.52  & 1.23 \\ \hline
100 & 1.55  & 1.25 \\ \hline
650 & 1.56  & 1.21 \\ \hline
\end{tabular}% }

    \end{subtable}
    \caption{Comparison of the quality of fits with different upper limits $Q^2_{\mathrm{up}}\; \mathrm{[GeV^2]} $ of the hard scale. }
    \label{table:Qup}
\end{table}

\begin{figure}[t]
    \resizebox{\textwidth}{!}{
        \includegraphics{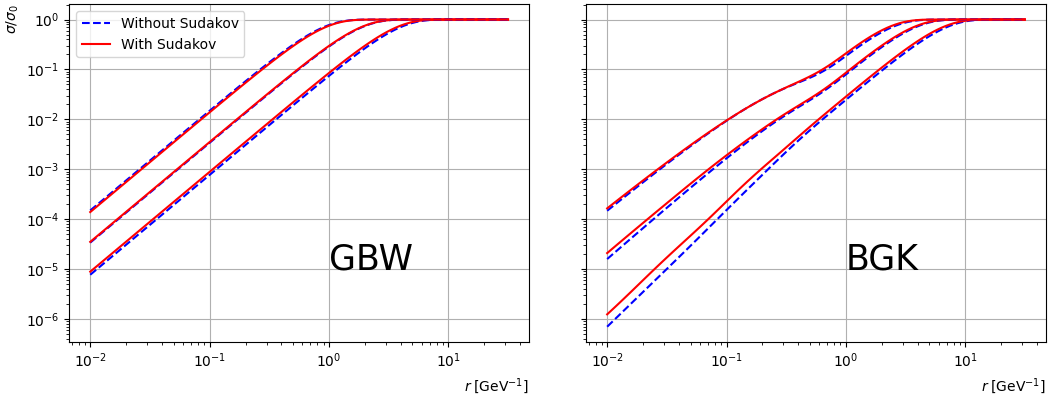}
    }
    \caption{The dipole cross section $\sigma_{\mathrm{dipole}}/\sigma_0$ at $x=10^{-2},\;10^{-4},\; 10^{-6}$ (from bottom to top).}
    \label{dipole}
\end{figure}

In this section, we proceed with numerical evaluation of the dipole cross section with the Sudakov factor, Eq.~(\ref{eq:sudakov}), incorporated. 
The values for $C_S$ and $\mu_{0S}^2$ in Eq.~(\ref{eq:newstar}) (The subscript $S$ was added to differentiate from those for the collinear gluon density.) 
for the Sudakov factor are fixed at $(2e^{-\gamma_\mathrm{E}})^2 \simeq 1.26 $ and 2 $\mathrm{GeV^2}$, respectively, and thus no additional fit parameters are needed. 
The choice of $\mu_{0S}^2$  is somewhat arbitrary, and if it was set higher, the effects of the non-perturbative Sudakov factor would become more relevant. We have checked that at sufficiently small value of $\mu_{0S}^2$, one can neglect the non-perturbative Sudakov factor (\textit{cf.} Tab.~\ref{table:Snp} in the Appendix). As it is preferable to have a model with fewer parameters, we chose to use the above value of $\mu_{0S}^2$ and omit the non-perturbative Sudakov factor.

As we fit the model to the data up to $Q^2=650\;\mathrm{GeV^2}$, the charm and bottom quarks are included.  
The masses for the $c$ and $b$ quarks are 1.3 GeV and 4.6 GeV, respectively, following the values used in Ref.~\cite{Golec-Biernat:2017lfv}. As for the light quarks, a massive case with $m_l=0.14$ GeV, the value used originally in Ref.~\cite{Golec-Biernat:1998zce}, and the massless cases were studied.  The Bjorken scaling variable~$x$ was modified as in Eq.~(\ref{eq:modx}). 
The parameters in the GBW and BGK models (and their improved versions), $\{\sigma_0, x_0, \lambda\}$ and  $\{\sigma_0, A_g, \lambda_g, C, \mu_0\}$, respectively, were fitted to the $F_2$ data from HERA~\cite{Abt:2017nkc} in the ranges $0.045\mathrm{GeV^2} \leq Q^2\leq 650 \mathrm{GeV^2}$ and $x \leq0.01$ using the MINUIT~\cite{minuit} package. 
In the rest of this section, GBW+Sud and BGK+Sud denote the respective improved versions of the saturation models. 

\begin{figure}[p]
    \includegraphics[width=0.5\textwidth]{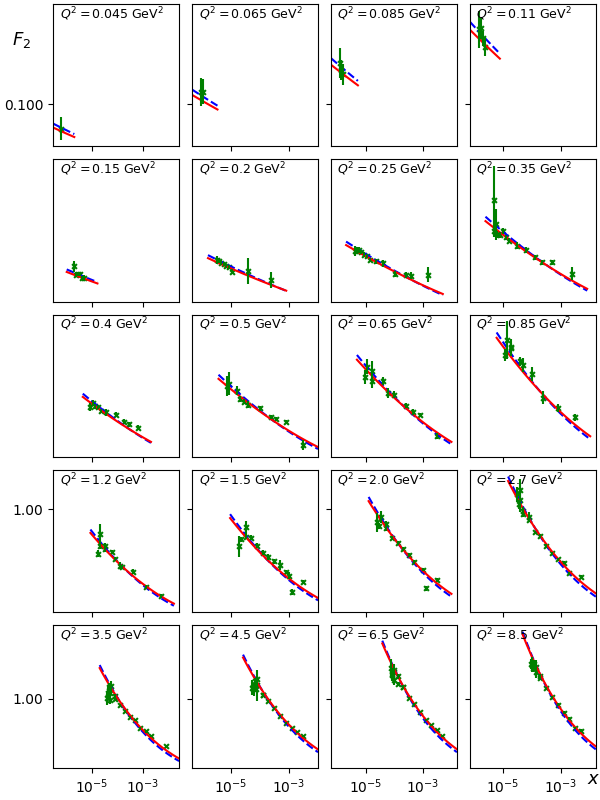}
    \includegraphics[width=0.5\textwidth]{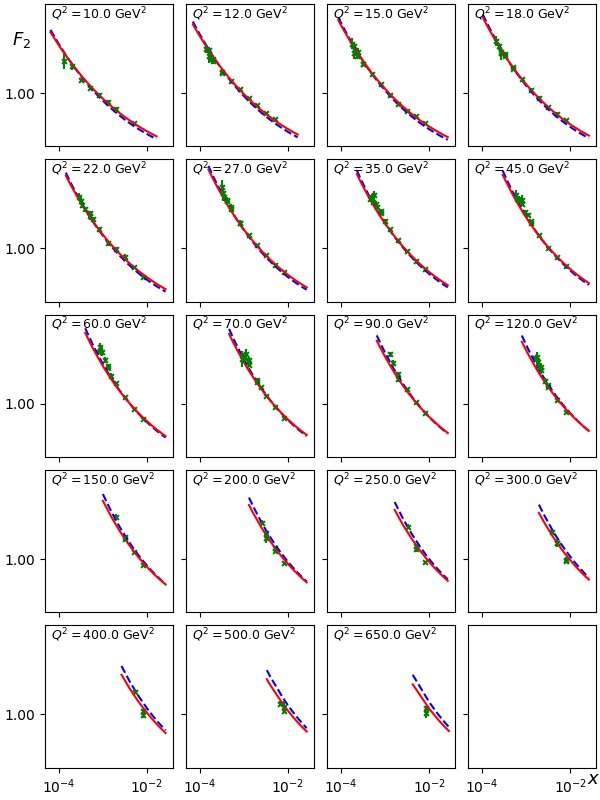}
    \includegraphics[width=\textwidth]{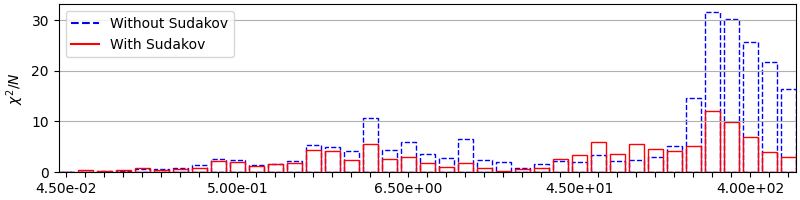}
    \caption{Top: $F_2$. GBW (dashed blue) and GBW+Sud (solid red) and HERA data (with error bars), with fixed values of $Q^2$. Bottom: $\chi^2/\text{(no. of points)}$ of GBW and GBW+Sud models. Significant improvement at large $Q^2$ is clearly visible.}
    \label{fig:gridGBW}
\end{figure}

\begin{figure}[p]
    \includegraphics[width=0.5\textwidth]{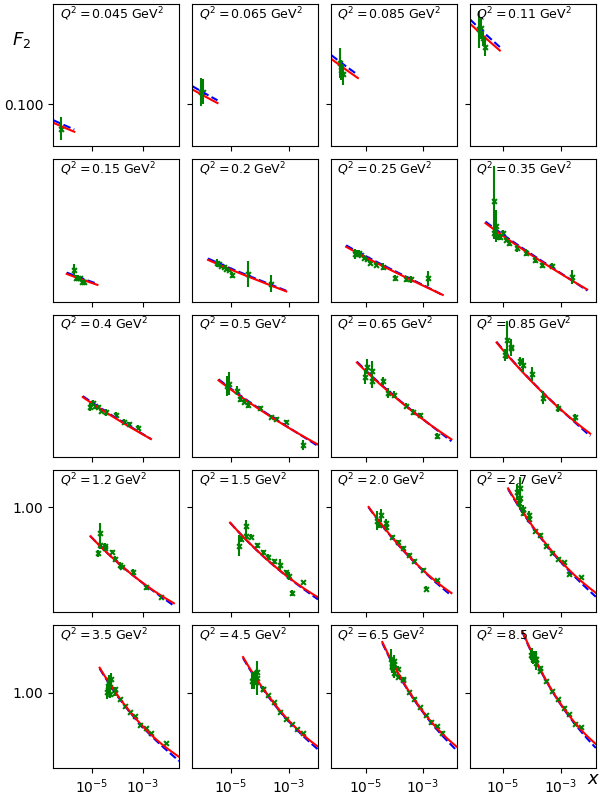}
    \includegraphics[width=0.5\textwidth]{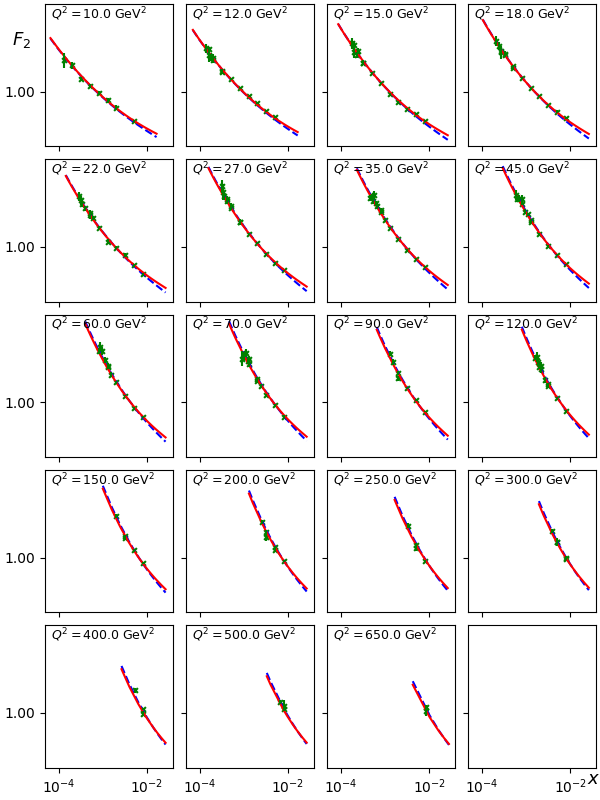}
    \includegraphics[width=\textwidth]{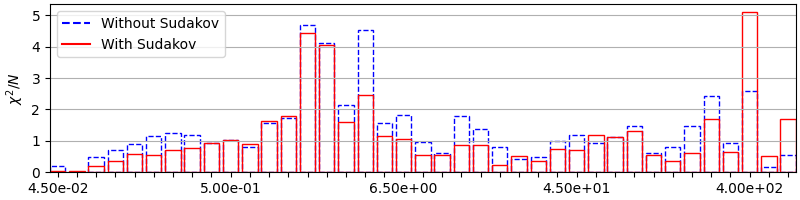}
    \caption{Top: $F_2$. BGK (dashed blue) and BGK+Sud (solid red) and HERA data (with error bars), with fixed values of $Q^2$. Bottom: $\chi^2/\text{(no. of points)}$ of BGK and BGK+Sud models.}
    \label{fig:gridBGK}
\end{figure}

The results are summarized in Tab.~\ref{table:fitresults}. As it was the case for Ref.~\cite{Golec-Biernat:2017lfv}, the best results are obtained with massless light quarks for both the GBW and BGK models. Hence, in this paper, we focus mainly on the massless cases. The fit results with massive light quarks are provided in the Appendix (Tab.~\ref{table:fullTable}).
As one can see in Tab.~\ref{table:fitresults}, differences in the parameters for the GBW and GBW+Sud models are moderate, yet the GBW+Sud fits show significant reduction of the $\chi^2$ value.
One noticeable change in the parameters of the BGK model is that $\lambda_g$ in the BGK+Sud model is negative while  the value for the BGK model is positive. As discussed in Ref.~\cite{Bartels:2002cj}, the negative value of  $\lambda_g$ indicates that the initial condition of the gluon is valence-like and vanishes in the limit $x\rightarrow0$. Hence, the rise with $1/x$ is solely due to the DGLAP evolution.

In Fig.~\ref{dipole}, log-log plots of the dipole cross section are presented. One can see that in the fits with the Sudakov factor, the small- to intermediate-$r$ region is changed slightly. This is due to the change in the fit parameters. 
One should, however, keep in mind that the main contribution of the dipole cross section for the structure function comes from the region $r\sim1/Q$, due to the convolution with the photon wave function. 

Tab.~\ref{table:Qup} shows significant improvement in the GBW model for high $Q_{\mathrm{up}}$, which is the upper limit of the selection of the data,  particularly for the 650~$\mathrm{GeV^2}$ case. This indicates that the Sudakov factor has indeed introduced additional hard scale dependence in the model. For the small-to-medium-$Q^2_{\mathrm{up}}$ cases, the GBW and BGK models exhibit similar patterns in improvements, and show better effects for a greater range of $Q^2$.
What is important here is that the Sudakov-improved models have better tolerance for a wide range of the photon virtuality, and the Sudakov-improved BGK model has almost no deterioration for having a wide range of $Q^2$.
\begin{figure}[p]
    \centering
    \includegraphics[width=\textwidth]{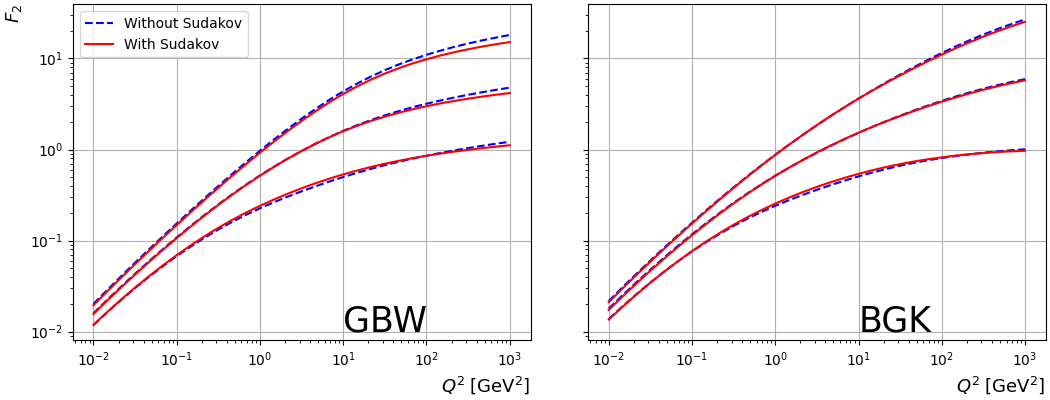}
    \caption{The structure function $F_2$. $x=10^{-2}, \;10^{-4},\; 10^{-6}$ (from bottom to top).}
    \label{F2}

    \vspace{2mm}

    \includegraphics[width=\textwidth]{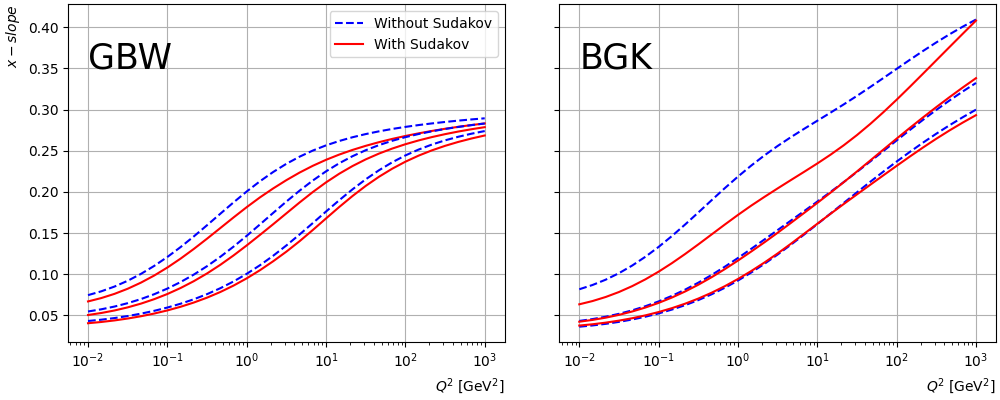}
    \caption{The $x$-slope of $F_2$,  $\lambda_\mathrm{eff}=-\frac{\partial \log F_2}{\log x}$. $x=10^{-2},\;10^{-4},\; 10^{-6}$ (from top to bottom). The slope is generally lowered, as can be seen also in Figs.~\ref{fig:gridGBW} and~\ref{fig:gridBGK}.}
    \label{slope}
    
    \vspace{2mm}

    \centering
    \includegraphics[width=\textwidth]{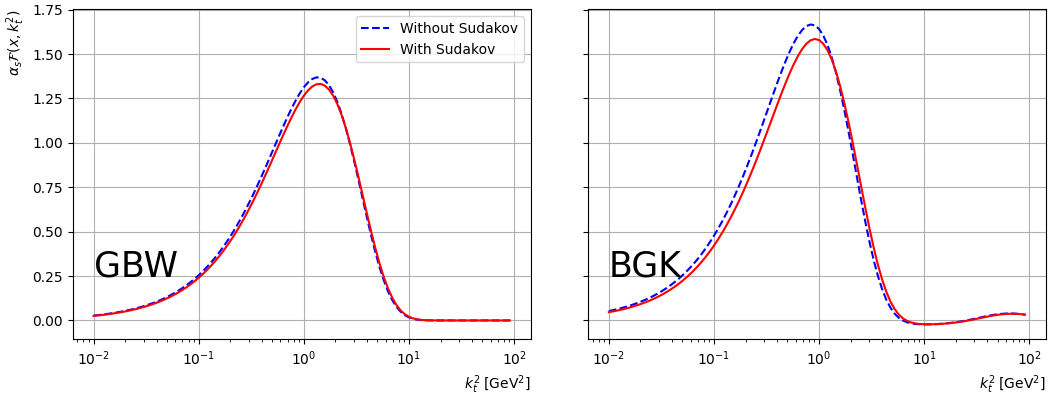}
    \caption{$\alpha_s\mathcal{F}(x,k_t^2)$ plotted for $x=10^{-4}$. }
    \label{fig:gluon}

\end{figure}

Figs.~\ref{fig:gridGBW} and~\ref{fig:gridBGK} show the $F_2$ function from our calculations plotted together with the HERA data~\cite{Abt:2017nkc}, where the bottom histograms depict the difference in $\chi^2$ value in each frame. 
The large-$Q^2$ region of the GBW model is modified significantly while the BGK model and the lower-$Q^2$ region of the GBW model show only slight change, mostly in the gradient. 
Fig.~\ref{fig:gridGBW} clearly shows that the highest improvement in the GBW model comes from the high-$Q^2$ region. 
As for the BGK model, shown in Fig.~\ref{fig:gridBGK}, the improvement comes more evenly from all values of $Q^2$, but the largest contribution comes from the medium-~to-low-$Q^2$ region, particularly the region around $\mu_0$ and $\mu_{0S}$. One can also see in the BGK model that, in the moderate-$x$ region, $F_2$ fits better to the data. This, as discussed in the introduction, is an expected effect of the Sudakov factor. 

The effect of the Sudakov factor in the structure function, $F_2$, is shown in Fig.~\ref{F2}. 
We see that for both the models, the Sudakov factor leads to a milder increase of $F_2$ with $1/x$, {\it i.e.} lowering the effective slope $\lambda_\mathrm{eff}=-\frac{\partial \log F_2}{\log x}$, {\it cf.} Fig~\ref{slope}, 
and lowering $F_2$ in the large-~and small-$Q^2$ region. This can also be seen in both Figs.~\ref{fig:gridGBW} and~\ref{fig:gridBGK}. 

\begin{figure}[p]

    \includegraphics[width=\textwidth]{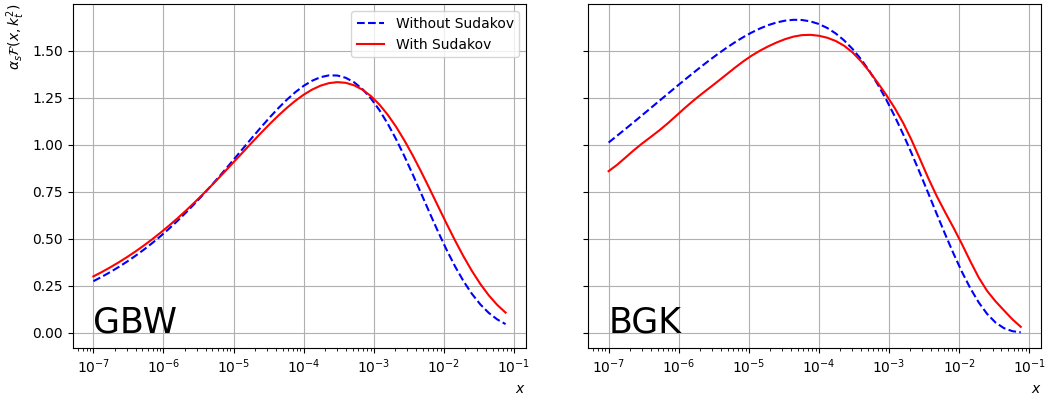}
    \caption{$\alpha_s\mathcal{F}(x,k_t^2)$ plotted for $k_t^2=1\;\mathrm{GeV^2}$. }
    \label{fig:gluon-x}
    
    \vspace{2mm}

    \includegraphics[width=\textwidth]{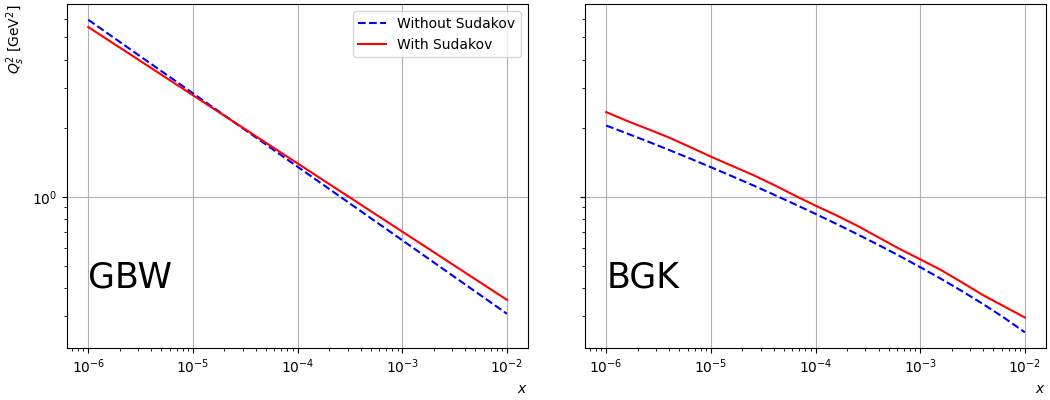}
    \caption{Saturation scale $Q_s^2(x)$.}
    \label{fig:critical}
    
    \vspace{2mm}

    \includegraphics[width=\textwidth]{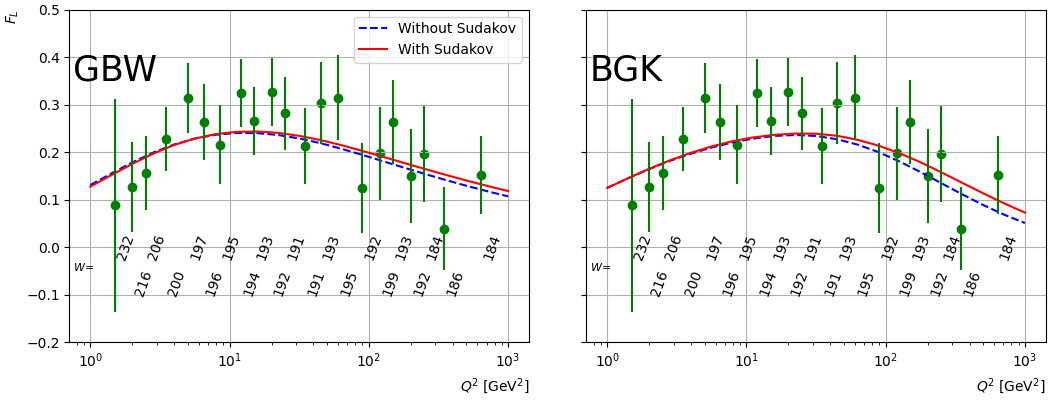}
    \caption{The longitudinal structure function, $F_L$, at fixed $W= 200$ GeV. The effects of the Sudakov factor is primarily visible in the large-$Q^2$ region, lifting up the line for both the GBW and BGK models. The data are from Ref.~\cite{H1:2013ktq}.}
    \label{fig:FL}
\end{figure}

The unintegrated gluon distributions obtained from the models are shown in Figs.~\ref{fig:gluon} and~\ref{fig:gluon-x} (for recent work on gluon density from the BGK model with updated fit of the model see \cite{Luszczak:2022fkf}).
Both models show changes in the distribution with the peak position moving slightly. Particularly interesting is the $x$ dependence of the dipole gluon density which shows the valence like behaviour in the saturated  region. This shape is a consequence of saturation and can be understood already with an example of the original GBW model.
\begin{equation}
    {\cal F}(x,k_t^2)=\frac{\sigma_0 N_c}{4\pi^2\alpha_s}\frac{k^2_t}{Q_s^2(x)}e^{-\frac{k^2}{Q_s^2(x)}}\,.
\end{equation}
For $k_t^2 \ll Q_s^2$ we get
\begin{equation}
    {\cal F}(x,k_t^2)\sim\frac{k_t^2}{Q_s^2(x)}
\end{equation}
which is a decreasing function of $x$. This behaviour is universal for dipole gluon densities with saturation and holds also for the BK equation \cite{Kutak:2006rn,Hentschinski:2022rsa}. From the same figure we see that the Sudakov factor modifies this behaviour.
To quantify the saturation, one introduces a line on the ($x,Q$) plane which separates the two regions where the saturation effects are important and  where the dipole cross section scales like $\sim r^2$.
The saturation scale, $Q_s(x)$, is an energy scale at a specific $x$. In other words, for a given value of $x$, saturation effects dominate below the scale $Q_s(x)$.
In order to study saturation in terms of the dipole cross section, one defines a saturation radius $R_0=2/Q_s$.  
Then, the condition for the saturation scale is defined as~\cite{Golec-Biernat:2006koa}
\begin{equation}
    \sigma(x,R_0)=a \sigma_0,
\end{equation}
where $a$ is a constant to be chosen.
Alternatively, one can use a definition of the saturation scale related to the emergence of maximum of the dipole gluon density~\cite{Kutak:2009zk}
\begin{equation}
    \frac{\partial \mathcal{F}(x,k_t^2)}{\partial k_t^2}=0.
\end{equation}

Comparisons of $x$-dependent saturation scale are presented in Fig.~\ref{fig:critical}. While the change in the GBW model is only in the slope, the saturation scale of the BGK model becomes higher for the whole range of $x$, implying the saturation effect to show at slightly higher scale than expected from the original model.

Finally, in Fig.~\ref{fig:FL} we present the longitudinal structure function $F_L$ together with the H1 data~\cite{Abt:2017nkc}. The main effects of the Sudakov factor occur in the large-$Q^2$ region, and, for both the GBW and BGK models, $F_L$ is higher when the Sudakov factor is present. 
Current experimental errors do not allow to favour any model shown in Fig.~\ref{fig:FL}.

%%%%%%%%%%%%%%%%%%%%%%%%%%%%%%%%%%%%%%%%%%%%%%%%%%%%%%%%%%%%%%%%%%%%%%%%%%%%%%%%%%%%%%%%%%%%%%%%%%%%%%
\section{Summary}

We obtained a general formula for incorporating the Sudakov factor into the dipole factorization. We then applied the formula to the well-known GBW and BGK saturation models.
The results of fitting such Sudakov-upgraded models to the HERA data in the range $0.045\mathrm{GeV} \leq Q^2 \leq 650\mathrm{GeV}$ and $x\leq 10^{-2}$ have shown improvements for both the GBW and BGK models and most importantly, the BGK model exhibits good behaviour over a wide range of $Q^2$. In terms of the dipole cross section of the GBW model, the changes are moderate, and reduces $x$-gradient slightly. 
Whereas, for the BGK model, while the small-$r$ region shows similar effect of reduced $x$-gradient, overall change is that the small- to medium-$r$ region of the dipole cross section is lifted. This difference in the effect on two models in the moderate-$r$ region is more clearly visible in the saturation scale; while the GBW changes its gradient over $x$, the BGK simply shifts $Q_s^2$ a little higher.   %%%%%%%%%%%%%%%%%%%%%%%%%%%%
\section*{Acknowledgements}
We are grateful to Krzysztof Golec-Biernat for numerous useful discussions and for careful reading of the manuscript. The project is partially supported by 
the European Union’s Horizon 2020 research and innovation program under grant agreement No. 824093.
TG and SS are partially supported by the Polish National Science Centre grant no. 2017/27/B/ST2/02004.
KK acknowledges the support of The Kosciuszko Foundation for the Academic year 22/23 for the
project ”Entropy of dense system of quarks and gluons”. Furthermore, KK  acknowledges the hospitality of the Nuclear Theory group at the BNL, where part of the  project was realized. 

\appendix
\section{Supplementary tables}

In this appendix we present some additional details concerning our fits of the saturation models.

\begin{table}[H]
    \begin{subtable}[H]{0.5\textwidth}
        \centering
        \resizebox{\textwidth}{!}{
            \begin{tabular}{|c|c|c|}
\hline
\footnotesize	$\mu_{0S}^2[\mathrm{GeV^2}]$ & \footnotesize $\mathrm{GBW+Sud_{pert}}$ & \footnotesize $\mathrm{GBW+Sud_{pert+np}}$\\\hline
 1 & 2.71 & 2.72 \\\hline
 2 & 2.66 & 2.67 \\\hline
 3 & 2.64 & 2.65 \\\hline
 4 & 2.64 & 2.64 \\\hline
 5 & 2.64 & 2.65 \\\hline
 \end{tabular} 

        }
        \hspace{1mm}
    \end{subtable}
    \begin{subtable}[H]{0.5\textwidth}
        \resizebox{\textwidth}{!}{
            \begin{tabular}{|c|c|c|}
\hline
\footnotesize	$\mu_{0S}^2[\mathrm{GeV^2}]$ & \footnotesize$\mathrm{BGK+Sud_{pert}}$ & \footnotesize $ \mathrm{BGK+Sud_{pert+np}}$\\\hline
 1 & 1.18 & 1.17 \\\hline
 2 & 1.21 & 1.17 \\\hline
 3 & 1.25 & 1.21 \\\hline
 4 & 1.29 & 1.21 \\\hline
 5 & 1.32 & 1.22 \\\hline
 \end{tabular} 

        }
    \end{subtable}
    \caption{$\chi^2/\mathrm{dof}$ of fits with and without the non-perturbative Sudakov factor, with 5 values of $\mu_{0S}^2$ for the Sudakov factor. It is clear that when $\mu_{0S}^2$ is set small, the effect of non-perturbative Sudakov factor is negligible. It is interesting to see that, for the BGK model, inclusion of the non-perturbative factor indeed weakens the dependence on the choice of $\mu_{0S}$.}
    \label{table:Snp}
    \vspace{3mm}
    \begin{subtable}[H]{\textwidth}
        \centering
        \resizebox{!}{0.075\textwidth}{\begin{tabular}{|c c|c|c|c|c|}
\hline
type & light quark mass $[\mathrm{GeV}]$&  $\sigma_0 \;[\mathrm{mb}]$ &  $x_0 (10^{-4})$ &  $\lambda$ &  $\chi^2/\mathrm{dof}$ 
\\\hline
$\mathrm{GBW}$& $m_l=0.14$ & 23.8 & 1.12 & 0.308 & 5.27 \\ \hline
$\mathrm{GBW+Sud}$&$m_l=0.14$ & 23.0 & 1.32 & 0.287 & 3.37 \\ \hline
$\mathrm{GBW}$& $m_l=0.0$ & 19.1 & 2.58 & 0.322 & 4.44 \\ \hline
$\mathrm{GBW+Sud}$& $m_l=0.0$ & 18.6 & 3.11 & 0.299 & 2.66 \\ \hline
\end{tabular} }

        \vspace{2mm}
    \end{subtable}

    \begin{subtable}[H]{\textwidth}
        \centering
        \resizebox{!}{0.075\textwidth}{\begin{tabular}{|c c|c|c|c|c|c|c|}
\hline
type&light quark mass $[\mathrm{GeV}]$&  $\sigma_0\; [\mathrm{mb}]$ &  $A_{\mathrm{g}}$ &  $\lambda_{\mathrm{g}}$ &  $C$ &  $\mu_{0}^2\;[\mathrm{GeV^2}]$ &  $\chi^2/\mathrm{dof}$ 
\\\hline
$\mathrm{BGK}$& $m_l=0.14$ & 33.1 & 1.33 & 0.0553 & 0.420 & 1.76 & 1.62 \\ \hline
$\mathrm{BGK+Sud}$& $m_l=0.14$& 30.7 & 6.28 & -0.380 & 0.677 & 3.09 & 1.27 \\ \hline
$\mathrm{BGK}$& $m_l=0.0$ & 23.3 & 1.18 & 0.0832 & 0.329 & 1.87 & 1.56 \\ \hline
$\mathrm{BGK+Sud}$& $m_l=0.0$ & 22.2 & 8.67 & -0.500 & 0.670 & 3.83 & 1.21 \\ \hline
\end{tabular}
}

    \end{subtable}
    \caption{ The parameters and $\chi^2$ per degrees of freedom  of the GBW and BGK models for the fit with the HERA data ($Q^2<650\;\mathrm{GeV^2}$). Two cases with massive light quarks and massless light quarks are fitted.}
    \label{table:fullTable}
\end{table}

%%%%%%%%%%%%%%%%%%%%%%%%%%%%%%%%%%%%%%%%%%%%%%%%%%%%%%%%
%\section{Weizs\"acker-Williams gluon density}
%One can also compute another type of unintegrated gluon densities, namely the  Weizs\"acke-Williams gluon distribution, which has a number density interpretation in CGC and can be directly accessed with dijet processes in DIS. It is also needed to obtain cross  section for dijets production in $p-p$ and $p-A$ collisions~\cite{Dominguez:2011wm,vanHameren:2021sqc}. This gluon density can be computed in terms of the dipole amplitude in the adjoint representation~\cite{vanHameren:2016ftb,Xiao:2017yya,Zheng:2014vka},

%\begin{equation}
%    \alpha_s \phi(x,k_t^2,Q^2)=\frac{C_F}{2 \pi^4}\int \frac{d^2\mathbf{r}}{r^2} e^{-i \mathbf{r}\cdot \mathbf{k}_t} e^{-S(r,Q^2)}\sigma_\mathrm{dipole, adj}(x,r),
%\end{equation}
%where 
%\begin{equation}
%   \sigma_\mathrm{dipole, adj}(x,r)=2 %\sigma_\mathrm{dipole}(x,r) - %\sigma_\mathrm{dipole}^2(x,r)/\sigma_0.    
%\end{equation}
%
%Fig.~\ref{fig:ww} shows $\alpha_s\phi(x,k_t^2,Q^2)$. The modifications due to the Sudakov factor are clearly visible and the dependence on $Q^2$ is pronounced.
%\begin{figure}[t]
%    \centering
%    \includegraphics[width=\textwidth]{Plots/ww-gluon.png}
%    \caption{Weizs\"acker-Williams gluon densities obtained from the respective models.}
%    \label{fig:ww}
%\end{figure}
%%%%%%%%%%%%%%%%%%%%%%%%%%%%%%%
\printbibliography
%%%%%%%%%%%%%%%%%%%%%%%%%%%%%%%
\end{document}